# BOTTOM-UP APPROACH TO SILICON NANOELECTRONICS

*Hiroshi Mizuta and Shunri Oda*

Quantum Nanoelectronics Research Center, Tokyo Institute of Technology
and SORST JST (Japan Science and Technology)

## ABSTRACT

This paper presents a brief review of our recent work investigating a novel bottom-up approach to realize silicon based nanoelectronics. We discuss fabrication technique, electronic properties and device applications of silicon nanodots as a building block for nanoscale silicon devices.

## 1. INTRODUCTION

Over the past few decades, the performance of silicon-based VLSI circuits has steadily been improved by scaling down device dimensions, and a nearly exponential growth of microelectronics capabilities has been achieved. However, maintaining this top-down miniaturization trend is getting exceedingly hard due to fundamental physical and technological limitations as well as of the economical limitation although the ITRS now predicts that the physical gate length of high-performance MOSFETs will reach sub-10-nm in 2016.. On the other hand, the use of organic molecules as a building block for nanoscale devices has attracted much attention since the precisely controlled nanostructures may be formed cheaply by utilizing self-assembly of molecules. This bottom-up technology can potentially overcome the inherent problem of the present silicon top-down technology. The conductivity of the organic molecular structures is, however, still much lower than those for silicon as the electron transport along the single molecule is basically governed by the hopping conduction.

Silicon nanodots (SiNDs) and nanowires (SiNWs) [1]-[4] may provide a solution to these issues by meeting the requirements both of bottom-up organization and superior carrier transport properties. As those silicon nanostructures can be formed on non-Si substrates such as glass and plastic, the Si-based bottom-up approach may lead to high-performance and large-area electronics. In addition, zero- and one-dimensional nature of electronic states in the individual SiNDs and SiNWs realizes new electronic and photonic properties which are not achieved with bulk silicon. Combining the bottom-up approach with the conventional top-down Si technologies enables us to explore silicon nano-, micro- and macro-electronics on the common technical footing.

In this paper we focus on the SiNDs in particular and present our recent studies on fabrication technique, unique electronic properties and novel device applications.

## 2. BOTTOM-UP SILICON NANOSTRUCTURE FABRICATION

### 2.1. Formation of nanocrystalline silicon dots

For fabricating SiNDs we have studied three different techniques. The first method is to use a very thin nanocrystalline (nc) Si film with the size of the grains down to a few nanometer. The nc-Si films can be formed either from an amorphous Si film with solid phase crystallization (SPC) or by using a very high frequency (VHF) plasma-enhanced chemical vapour deposition (PECVD) at a low temperature [5]. In the SPC films the individual grains are usually columnar shaped, and the grain boundaries (GBs) between adjacent grains contain carrier trap states due to dangling bonds. On the other hand, in the PECVD films, the individual grains are more spherical, and the GBs are formed by a-Si:H layers between grains. The second approach is to use porous Si [6] formed by using photoanodization of the Si substrate. The surface of the SiND islands formed in the substrate can be oxidised selectively by electrochemical oxidation. Formation of a linear chain of SiND islands with a diameter as small as 5 nm has been observed [7].

The most recent approach is a VHF plasma enhanced deposition of silane ($SiH_4$) with a hydrogen gas pulse sequence [8]. This technique facilitates in separating the nucleation and crystal growth process and enables to fabricate single crystalline SiNDs (Fig. 1) with diameter less than 10 nm. The nucleated SiNDs are grown in the $SiH_4$ plasma in the intervals between $H_2$ gas pulses, and dot diameter is therefore determined by the growth time of the SiNDs in the plasma cell. Using this method, we fabricated SiNDs with diameters of around 8 nm and a very narrow size spread (±1 nm). Using a self-limiting oxidation process, it is possible to reduce the dot size further down to 4 nm. The interparticle tunnel barriers can also be formed by in-situ oxidation or nitridation in a controlled manner.

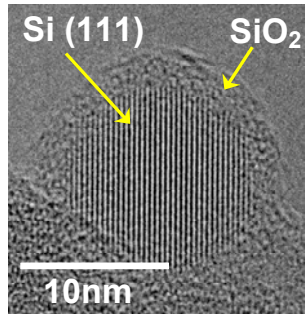

Fig. 1  Single crystalline silicon nanodot covered with thin $SiO_2$

## 2.2. Combining bottom-up and top-down approaches for fabricating nanoscale device structures

Integrating the fabricated SiNDs, either over a large area or in a local area, is a very challenging issue. Various techniques are currently under investigation, for example, the dispersion solution drop & evaporation method [9], the Langmuir Blodgett (LB) method and the nano template method, We also examined a novel method of fabricating nanoscale devices by conducting the self-assembly of the SiNDs on patterned nano electrodes [10][11]. For preparing the SiND dispersion solution, the substrates with deposited SiNDs were soaked into the solvents such as methanol or isopropanol, and the ultra sonic treatment was applied for a few tens seconds. The SiND solution was then condensed by evaporating the solvent a fraction.

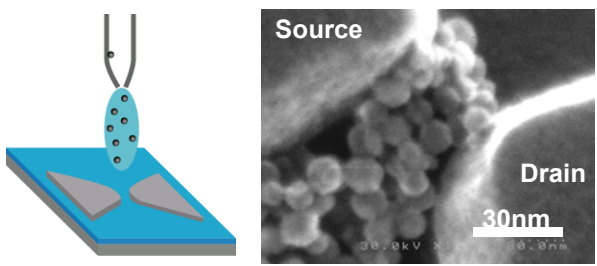

Fig. 2  SiNDs assembled on the nano electrodes by using the dispersion solution method

The nanoelectrodes were fabricated by using the electron beam lithography on the heavily-doped n-type silicon on insulator (SOI) substrate with thickness of about 50nm and the buried oxide (BOX) thickness of 200nm. This structure gives a good areal contrast of hydrophobic (Si) and hydrophilic ($SiO_2$) surfaces and works as a template for the following SiND assembly process. We used a drop and evaporation technique [9] to assemble the SiNDs from the dispersion solution by using the lateral capillary meniscus force, which operates at the point where the three phases of the liquid, air and SiND meet. Figure 2 shows the SiNDs assembled on the SOI substrate with a patterned nanogap of about 35 nm. We observed that SiNDs remained only in the $SiO_2$ regions and were assembled near the nano gap, resulting in a SiND channel between the electrodes. This trend was observed in common over the large area. This method may be useful to fabricate the SiND based nanoscale transistors, and we believe that it is possible to form a channel with few SiNDs or even a single dot by reducing the density of SiNDs and optimizing evaporation conditions.

## 3. ELECTRON TRANSPORT PROPERTIES OF SILICON NANODOTS

### 3.1. Resonant tunneling via single Si nanodot

Strong quantum confinement effect in a SiND is the key to realize Si based resonant tunneling devices. We characterized tunneling current through a single SiND by using the contact mode AFM [12]. The SiNDs were sparsely distributed on the n-type [100] Si substrate with a surface density of. $1.4 \times 10^8$ $cm^{-2}$. For the cantilever we used the silicon tip coated with gold. We performed contact-AFM scanning with 1 x 1 $\mu m^2$ scan area to generate the topography image. The AFM measurement was performed by selecting a single SiND from the topographical view.

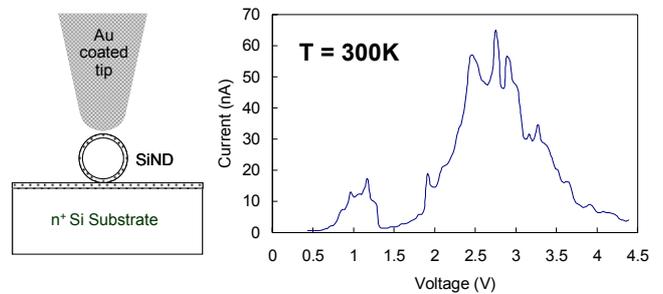

Fig. 3  Resonant tunnelling current via a single SiND measured at room temperature by using contact mode AFM.

I-V characteristics were measured for a single SiND at room temperature [11]. Figure 3 shows the typical I-V curve observed when the sample was positively biased from 0 to 4.5 V. Two current peaks were clearly observed: the first peak around 1 V and the second one around 2.5 V. The largest peak-to-valley (P/V) current ratio observed is 17 for the 1st resonance. As SiND based resonant tunneling structures can be integrated with conventional MOSFETs, the observed multiple negative differential conductance characteristics may be used for making Si based novel functional devices such as tunnel-based SRAM and resonant tunneling transistors.

### 3.2. Coulomb blockade and electron interaction in coupled double Si nanodots

Charge quantization and Coulomb blockade have also been studied intensively for SiNDs. Coulomb oscillation of tunneling current has been observed at room temperature by using point-contact single electron transistors (PC-SETs) with very few SiNDs in the channel. Electrostatic and coherent coupling effects have also been studied recently for strongly-coupled double SiNDs by using PC-SETs fabricated on a nc-Si thin film. The PC-SETs with a very small channel with 30 nm × 30 nm in lateral dimensions were formed on a 40 nm thick nc-Si film with lateral grain size of 20 – 25 nm. The electrostatic potential on the grains are controlled via the bias applied to two side gates.

The PC-SETs exhibited delocalisation of the electron wavefunctions over the coupled double dots via a very thin tunnel barrier. A plot of the device conductance at 4.2 K as a function of the two side gate voltages shows single-electron conductance peaks which partially form an electron stability diagram for two charging dots (Fig. 4).

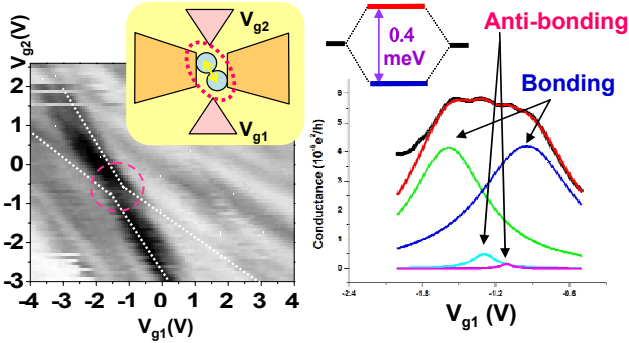

Fig. 4 Coherent electron coupling observed for strongly-coupled double SiNDs

The peak lines in this plot (white dotted lines) show strong splitting (a dotted circle) caused by electrostatic interactions when the energy levels in the two dots are in resonance [13]. In this strong coupling region, we observed that the characteristics are decomposed into four Lorentzian peaks – two main peaks with two small peaks (Fig. 4; right figure) [14]. This is attributed to the tunnel coupling across two adjacent double dots, resulting in bonding- and anti-bonding-like resonance peaks. Tunnel splitting obtained from the peak separation is about 0.4 meV, which is from a few time to an order-of-magnitude larger than the value reported previously in GaAs/AlGaAs quantum dots. Such quasi-molecular states may be used to realize a Si-based charge qubit.

### 3.3. Phononic band formation in Si nanodot array

Silicon and $SiO_2$, the key players in the present VLSIs, now combine in a different way to offer new functional applications in electronics and mechanics. Electron transport properties of the SiNDs interconnected with thin oxide layers have recently attracted growing attention due to the experimental observation of ballistic electron emission [15] as well as the theoretical study of phonon depletion [16]. The electronic and phononic states have recently been investigated for a one-dimensional array of SiNDs interconnected with thin oxide layers as shown in Fig. 5. This acoustic heterostructure has wide variety of interesting features such as phonon bandgap and phonon confinement. Changing the thickness of the oxides controls the energy range of the bandgap. It is also interesting to note that the energy dispersion for high-energy phonons is flat. Such phonons are confined in SiNDs.

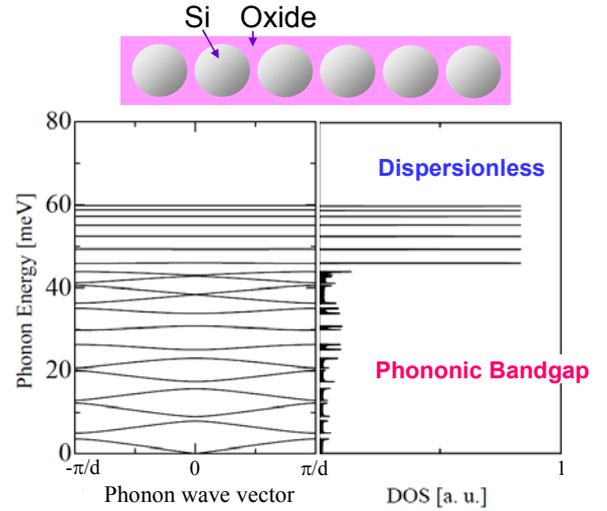

Fig. 5 Concept of SiND / SiO2 acoustic heterostructure and calculated phonon energy dispersion (left) and associated phonon DOS (right)

Another interesting feature of this structure is the reduction of the electron-phonon scattering potential, which is written as $H_{el-ph}(x) = D_{aco} \partial S(x)/\partial x$ where $S(x)$ and $D_{aco}$ denote phonon wave function and the coupling constant, respectively. The first derivative of $S(x)$ is also know as the strain. Figure 6 compares the strain in the acoustic heterostructure (circles with solid line) and conventional Si (broken line). Note that the oxide layers 'absorb' the strain from the SiNDs [17]. This is reasonable because the oxide layers are 'softer' than SiNDs (the Young's modulus of Si and oxide are 180GPa and 70GPa, respectively). As the coupling constant $D_{aco}$ in the oxide is smaller than that in Si, the strain absorption effect reduces the scattering potential over the entire region compared to that of Si. It was shown theoretically [18][19] that the electron energy loss rate is significantly suppressed at the vicinity of the miniband bottom energy.

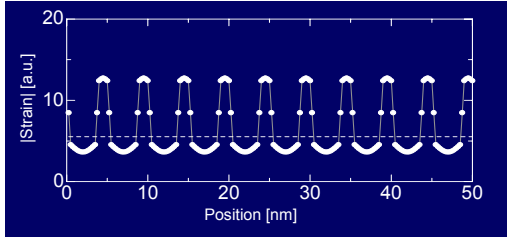

Fig. 6 Phonon accumulation in SiO$_2$ and associated phonon depletion in SiNDs obtained for acoustic phonon modes

### 4. NOVEL FUNCTIONAL DEVICE APPLICATIONS BASED ON SILICON NANODOTS

On the basis of the new properties shown in Section 3, we have pursued various device applications such as ballistic electron emission devices, light-emitting devices, single-electron transistors [20], SiND memory with a long data retention time [21], and a nonvolatile nano electromechanical system (NEMS) memory. Two unique device applications are introduced in the following:

#### 4.1. Ballistic electron emission device

As one of the promising applications of the SiNDs we investigated the SiND based electron surface emitter [22]. Unique phonon properties of the array of SiNDs covered with SiO$_2$ shown in Section 3.3 lead to ballistic electron emission phenomenon. An electron emitter device was fabricated by using a 150-nm-thick layer of Si nanodots deposited on n+-Si substrate and a very thin Au top electrode (Fig. 7). When the voltage is applied across the SiND layer, only electrons with energy higher than the Au work function $W_{Au}$ of approximately 5 eV are emitted into vacuum through the top electrode. Both diode and emission currents were measured as a function of the extraction voltage $V_{ex}$, and the emission efficiency can be as large as 5%.

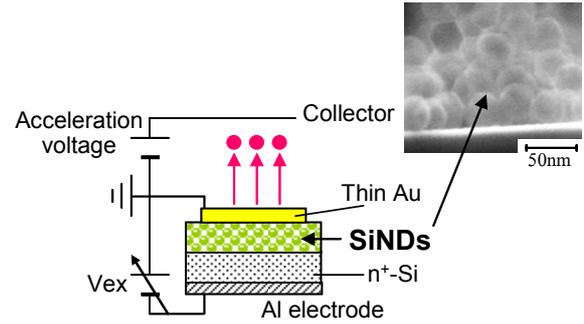

Fig. 7 Ballistic electron emitter based on stacked SiNDs

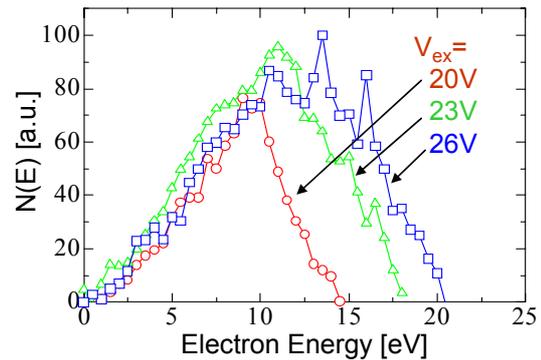

Fig. 8 Energy distribution of electrons emitted from the SiND layer

We also investigated the energy distribution of emitted electron by using a conventional ac-retarding-field analyzer. As shown in Fig. 8, we found that the energy distribution of the emitted electrons is non-Maxwellian in contrast with those observed for conventional cold emitters. Maximum electron energy agrees approximately with $eV_{ex} - W_{Au}$, and the peak energy varies in proportion to $V_{ex}$. These results show that fractional electrons travel throughout the SiND layer in a quasi-ballistic manner and emit into vacuum.

#### 4.2. Nanoelectromechanical memory device

A new nonvolatile memory concept has recently been proposed, based on the bistable operation of the sub-μm-long NEMS structure, combined with the SiNDs (Fig. 9; above) [23]. It features a suspended SiO$_2$ beam which incorporates the SiNDs as single-electron storage (Fig. 9; below). The beam may be moved via the gate electric field, and its positional displacement is sensed via a change in the drain current of the MOSFET underneath.

A free-standing SiO$_2$ single beam was first fabricated using a Si undercut etching technique. Most fabricated samples show the beams bent upwards as a result of mechanical stress, stored in the $SiO_2$ being released. The mechanical bistability of the beam was demonstrated by using the nano-indentor type loading system. Also, a top gate electrode has been fabricated successfully above the $SiO_2$ (Fig. 10).

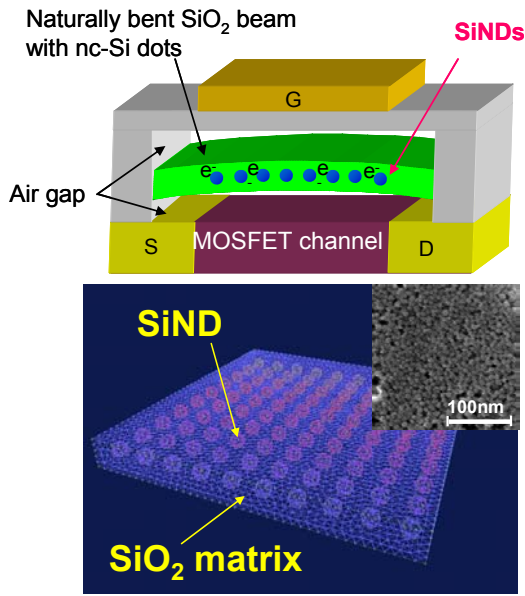

Fig.9 Schematic illustration of nanoelectromechanical memory structure (above) and a floating gate incorporating SiND array (below).

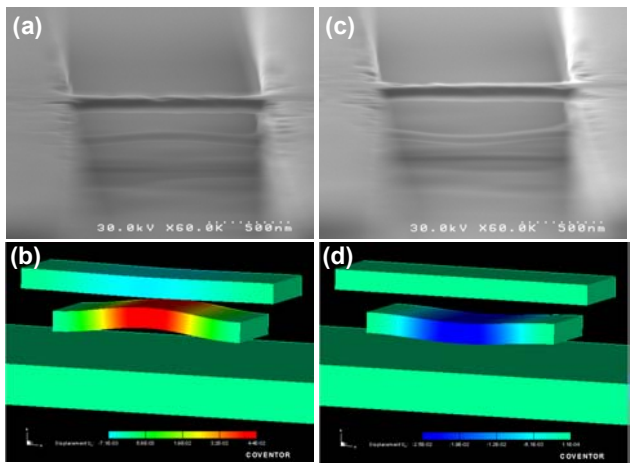

Fig.10 Fabricated and simulated NEMS beams with a top gate electrode at its bistable states.

The switching speed and power of the beam have been studied intensively by using a 3D FEM simulation [24]. The switching speed increases approximately in inverse proportion to the beam length L and goes beyond 1 GHz in the sub-µm regime. The switching voltage of less than 15 V has recently been demonstrated for an optimized beam structure along with the scaling properties of the beam. In contrast with other emerging nonvolatile memories such as MRAM or PCRAM, the NEMS memory can still be manufactured within the conventional Si technologies. With further optimization, it may exhibit excellent characteristics comparable to them.

## 5. CONCLUSION

We have studied SiNDs as a promising building block for Si nanoelectronics. SiNDs with diameter as small as 5 nm can be fabricated by using VHF plasma CVD, and interface tunnel layer thickness of 1 nm were achieved. In addition, the SiND assembly technique was successfully combined with the nanofabrication technique to build nanoscale transistors with SiNDs as a channel. A variety of new functions were found for SiNDs: ballistic electron emission, single-electron charging effects, and quantum-mechanical coupling between two adjacent dots, which have never been observed before in bulk silicon. Based on these unique material properties, we have explored various novel device applications such as a ballistic surface electron emitter and a high-speed and nonvolatile nanoelectromechanical memory. An innovative fusion of top-down and bottom-up silicon technologies may lead to a high-performance & low-cost material platform for macro-, micro- and nanoelectronics.

## ACKNOWLEDGMENT

The authors are very grateful to Dr. Y. Tsuchiya, Dr. K. Usami, Dr. M. Khalafalla, Dr. S. Huang, Dr. K. Nishiguchi (now of NTT Basic Res. Lab.), Mr. K. Takai (now of Toshiba Corp.), Mr. T. Nakatsukasa (now of Toppan Printing Co. Ltd.), Mr. A. Surawijaya, Mr. N. Momo, Mr. T. Nagami, Mr. A. Tanaka, Mr. G. Yamahata, Mr. J. Ogi of Tokyo Institute of Technology, Dr. Z.A.K. Durrani of Cambridge University, Dr. S. Uno of Nagoya University and Prof. N. Mori of Osaka University for their valuable technical contributions, and Dr. S. Saito, Dr. T. Arai of Central Research Laboratory, Hitachi Ltd., Dr. T. Shimada of Quantum 14 Co. Ltd., Prof. K. Nakazato of Nagoya University, Prof. N. Koshida of Tokyo University of A&T and Dr. T. A. Armour of Nottingham University for very useful discussions.